\documentclass[]{interact}

\usepackage{epstopdf}
\usepackage[caption=false]{subfig}

\usepackage[numbers,sort&compress]{natbib}
\usepackage{graphicx,amssymb,amsmath}
\usepackage{dcolumn}
\usepackage{bm}
\usepackage[colorlinks=true,citecolor=blue,linkcolor=blue,urlcolor=blue]{hyperref}
\usepackage{color}
\usepackage{xspace}

\bibpunct[, ]{[}{]}{,}{n}{,}{,}
\makeatletter
\def\NAT@def@citea{\def\@citea{\NAT@separator}}
\makeatother

\theoremstyle{plain}

\theoremstyle{definition}

\theoremstyle{remark}

\newcommand{\CFC}{CsFeCl$_{3}$\xspace}

\begin{document}


\title{Zr-based bulk metallic glass clamp cell for high-pressure inelastic neutron scattering}

\author{
\name{
S.~Hayashida\textsuperscript{a}\thanks{CONTACT S.~Hayashida. Email: s{\_}hayashida@cross.or.jp}, 
T.~Wada\textsuperscript{b},
M.~Ishikado\textsuperscript{a},
K.~Munakata\textsuperscript{a},
K.~Iida\textsuperscript{a},
K.~Kamazawa\textsuperscript{a},
R.~Kajimoto\textsuperscript{c},
Y.~Inamura\textsuperscript{c},
M.~Nakamura\textsuperscript{c},
K.~Iwasa\textsuperscript{d},
K.~Ohoyama\textsuperscript{d},
H.~Kato\textsuperscript{b},
H.~Kira\textsuperscript{a},
M.~Matsuura\textsuperscript{a},
and Y.~Uwatoko\textsuperscript{a,e,f}}
\affil{
\textsuperscript{a}Neutron Science and Technology Center, Comprehensive Research Organization for Science and Society (CROSS), Tokai, Ibaraki 319-1106, Japan; 
\textsuperscript{b}Institute for Materials Research, Tohoku University, Katahira 2-1-1, Aobaku, Sendai 980-8577, Japan;
\textsuperscript{c}Materials and Life Science Division, J-PARC Center, Japan Atomic Energy Agency (JAEA), Tokai, Ibaraki 319-1195, Japan;
\textsuperscript{d}Research and Education Center for Atomic Sciences and Graduate School of Science and Engineering, Ibaraki University, Tokai, Ibaraki 319-1106, Japan;
\textsuperscript{e}Faculty of Science and Engineering, Tokyo City University, 1-28-1 Tamazutsumi, Setagaya, Tokyo 158-8557, Japan;
\textsuperscript{f}Department of Advanced Materials Science, The University of Tokyo, Kashiwa, Chiba 277-8561, Japan}
}

\maketitle

\begin{abstract}
We report the fabrication and characterization of a Zr-based bulk metallic glass (Zr-BMG) clamp cell designed for high-pressure inelastic neutron scattering (INS) measurements.
The INS spectra of the empty cell exhibit broad and featureless backgrounds, reflecting the amorphous structure of the Zr-BMG. 
Test measurements using a reference sample, \CFC, confirm that the neutron transmission of the Zr-BMG cell is significantly higher than that of a conventional monobloc CuBe clamp cell.
These results demonstrate that the Zr-BMG clamp cell provides both enhanced neutron transparency and a clean background profile, thereby advancing high-pressure INS studies.
\end{abstract}

\begin{keywords}
Neutron scattering; bulk metallic glass; piston-cylinder clamp cell
\end{keywords}

\section{Introduction}
High-pressure inelastic neutron scattering (INS) is a powerful spectroscopic technique~\cite{Somenkov2005,Klotz2012,Fogh2025} that directly probes elementary excitations driven by pressures across a broad range of momentum and energy transfers.
This capability provides deep insight into fundamental interactions governing pressure-induced physical phenomena.
However, implementing INS under high pressure presents significant technical challenges.
In general, INS measurements require relatively large sample volumes because of their weak signals.
This in turn necessitates the use of bulky pressure cells, such as Paris–Edinburgh (PE) presses~\cite{Besson1992,Besson1995,Klotz2004} and McWhan cells~\cite{McWhan1974}.
Although successful INS studies using these cells were reported~\cite{HattoriPRB2022,FoghPRL2024,Zayed2017}, their large size imposes major limitations, particularly in terms of neutron transmission and background scattering.
Herein, piston-cylinder clamp cells have been developed for high-pressure INS experiments~\cite{Wang2011,Podlesnyak2018,Hattori2022,Yuan2022}, and widely used to date~\cite{PerrenPRB2015,HayashidaSciAdv2019,Hong2022}.
The clamp cells offer a simple design, moderate pressure range ($< 3$~GPa), and a relatively large sample capacity.

Despite their widespread use, conventional cell materials such as CuBe and NiCrAl alloys inherently degrade data quality due to their low neutron transmission and substantial background scattering~\cite{Kibble2019}.
For example, CuBe alloys exhibit a neutron transmission of only 36{\%} at a neutron energy of 10~meV for a 1~cm path length~\cite{Hattori2022}, accompanied by complex background features arising mainly from phonon modes in the inelastic scattering regime~\cite{Kibble2019}. 
NiCrAl alloys provide even lower neutron transmission, along with the phonon background.
These issues hinder advancing high-pressure INS studies and strongly motivate the further development of pressure cells with improved neutron transparency and background scattering.

A major breakthrough has been achieved with the introduction of Zr-based bulk metallic glass (Zr-BMG)~\cite{Inoue2000} as a pressure-cell material~\cite{Komatsu2015}.
Zr-BMG exhibits exceptional mechanical properties, including a high tensile strength of $\sim$1800~MPa combined with a Young's modulus of $\sim$80~GPa, and a large elastic elongation of 2{\%}~\cite{Inoue2000}.
For neutron scattering experiments, it also offers superior neutron transmission (67{\%} at 10 meV with 1~cm thickness~\cite{Komatsu2015}), exceeding that of conventional Cu- and Ni-based alloys.
Moreover, Zr-BMG is nonmagnetic, comparable to CuBe alloys~\cite{Komatsu2015}.
Most notably, its amorphous structure eliminates sharp Bragg reflections and acoustic phonons, resulting in a cleaner background in neutron scattering measurements.
Indeed, Zr-BMG piston-cylinder clamp cells have already been demonstrated to be highly effective for neutron diffraction experiments~\cite{Matsuda2016,Chi2016,Matsuda2018,Yamashita2020,Dissanayake2022,Matsuda2025}.
Thus, applying a Zr-BMG cell to INS experiments could offer significant advantages for investigating elementary excitations under pressure. 

In this study, we present the development of a Zr-BMG piston-cylinder clamp cell specifically designed for INS measurements.
Background and test-sample measurements using the designed pressure cell demonstrate its superior performance compared with the conventional CuBe-based clamp cells.

\section{Experimental details}\label{sec:experimental_details}
\begin{figure}[t]
\centering
\includegraphics[scale=1]{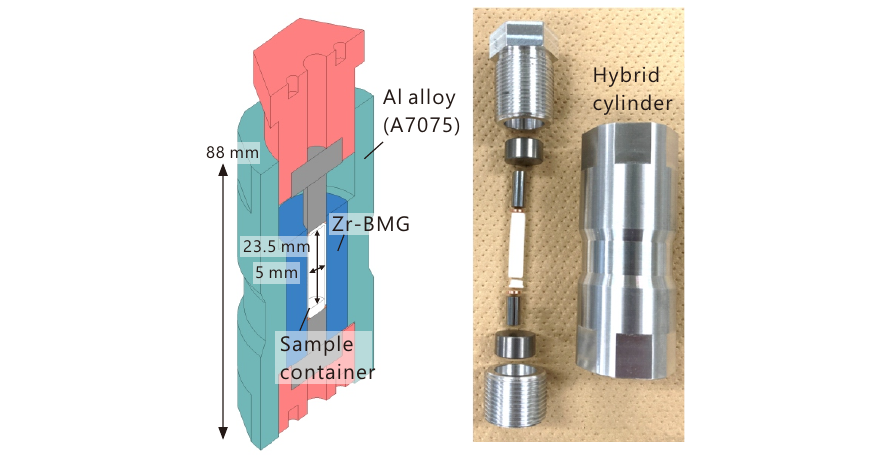}
\caption{Schematic view and photograph of the full assembled hybrid cylinder clamp cell used in the present study. The blue, green and gray components correspond to the Zr-BMG inner sleeve, the aluminum alloy (A7075) outer body, and the tungsten carbide piston and support spacers, respectively. The red components are the locking nuts fabricated from either aluminum alloy or CuBe alloy.}
\label{fig:pressure_cell}
\end{figure}
A schematic of the piston-cylinder clamp cell is shown in Fig.~\ref{fig:pressure_cell}. 
Since the synthesizable size of Zr-BMG rods is limited by the critical diameter for glass formation, a hybrid-cylinder cell was adopted instead of a monobloc cell. 
The inner sleeve is fabricated from Zr-BMG (composition Zr$_{55}$Al$_{10}$Ni$_{5}$Cu$_{30}$), while the outer body is made of an aluminum alloy (A7075) for minimizing neutron absorption.
The Zr-BMG rods with a diameter of 20~mm were synthesized using an arc-melting casting furnace.
For assembly, the interface between the inner sleeve and the outer body was slightly tapered to enable shrink fitting and ensure mechanical stability under load.
The inner diameter of the Zr-BMG cylinder is 6~mm.
The wall ratio of the outer/inner radii of the Zr-BMG cylinder is 3.3, potentially yielding a pressure of approximately 2~GPa~\cite{Komatsu2015}.
In practice, the cell was confirmed to remain stable under applied pressures above 1~GPa (Appendix~\ref{sec:HQR}).
The cell was also verified to reach temperatures of 300~mK ($^{3}$He refrigerator) and 170~mK ($^{3}$He-$^{4}$He dilution refrigerator).
In addition, we confirmed that the magnetic susceptibility of the synthesized Zr-BMG is comparable to that reported previously~\cite{Komatsu2015}.

Samples are enclosed in a polytetrafluoroethylene (PTFE) capsule with an inner diameter of 5~mm and a length of 23.5~mm.
Taking into account capsule compression under pressure, the effective sample volume is approximately 190~mm$^{3}$, corresponding to sample dimensions of 4.5~mm in diameter and 12~mm in length.
To suppress background scattering during neutron scattering measurements, cadmium sheets were placed over the top and bottom edges of the pressure cell.

The INS measurements were carried out on the time-of-flight neutron spectrometer 4SEASONS~\cite{4SEASONS} installed at Materials and Life Science Experimental Facility (MLF), Japan Proton Accelerator Research Complex (J-PARC), Japan. 
A Fermi chopper frequency of 150~Hz was employed, providing incident neutron energies of $E_{\rm i}=2.96$, 10.2, 20.0, and 55.6~meV, simultaneously.
The detector arrays cover scattering angles ($2\theta$) of $-35^{\circ}\leq 2\theta \leq 130^{\circ}$ horizontally and $-25^{\circ}\leq 2\theta \leq 27^{\circ}$ vertically.
An oscillating radial collimator was installed to reduce background scattering from the sample environment. 
The collimator has inner and outer radii of 210 and 400~mm from the sample center, respectively, with an angular separation of 2.5$^{\circ}$~\cite{Nakamura2015,Nakamura2018}.

A closed-cycle cryostat was used to achieve temperatures in the range of $6\leq T \leq 300$~K.
A Zr-BMG rod with a diameter of 15~mm and a length exceeding 40~mm was measured at room temperature. 
In addition, the empty Zr-BMG hybrid cell, without a pressure-transmitting medium or sample, was measured at 6~K and room temperature.
Both the Zr-BMG rod and the empty cell were mounted in the cryostat with their longitudinal axes perpendicular to the horizontal plane.
Test measurements were conducted using a \CFC single-crystal sample, grown by the vertical Bridgman method. The details of the synthesis procedure are described in Ref.~\cite{KuritaPRB2016}.
A crystal with dimensions of approximately $2.5\times3\times8$~mm$^{3}$ (mass: 260~mg) was used in this study.
For the INS measurements, the scattering plane was chosen such that the crystallographic $ab$-plane was horizontal.
Measurements were first performed with the sample mounted on an aluminum plate, and subsequently repeated with the same sample placed inside the pressure cell filled with deuterated glycerol as the pressure-transmitting medium.
In both configurations, the sample was cooled to 6~K and rotated over 62$^{\circ}$ in 1$^{\circ}$ increment, with each scan lasting 11~minutes.

Data reduction for all collected INS spectra was performed using Utsusemi software~\cite{Utsusemi}, and INS data were analyzed using HORACE software~\cite{Horace}.

\section{Results and discussion}
\subsection{INS on Zr-BMG rod}
\label{sec:BMG}

\begin{figure}[!h]
\centering
\includegraphics[scale=1]{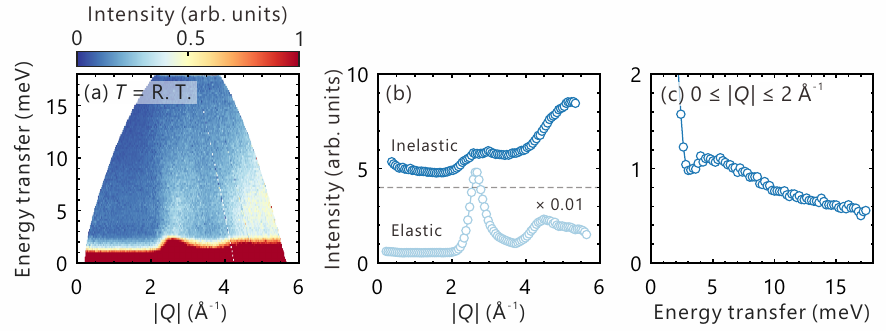}
\caption{(a) False-color plot of the INS spectrum of the Zr-BMG rod measured with $E_{\rm i}=20.0$~meV at room temperature (R.T.). (b) Momentum-transfer profiles of the spectra at the elastic line integrated over $-0.5 \leq E \leq 0.5$~meV and the inelastic regime integrated over $4 \leq E \leq 16$~meV. For visibility, the elastic profile is scaled by a factor of 0.01, and the inelastic profile is vertically offset by 4. (c) Energy-transfer profile of the INS intensity integrated below 2~{\AA}$^{-1}$.}
\label{fig:INS_BMG}
\end{figure}

The INS spectrum of the Zr-BMG rod is shown in Fig.~\ref{fig:INS_BMG}(a).
In contrast to sharp acoustic phonon features observed in CuBe, NiCrAl, and aluminum alloys~\cite{Kibble2019}, the INS spectrum of the Zr-BMG is notably broad, characteristic of an amorphous structure.
At the elastic line, a diffuse peak appears near $Q=2.6$~{\AA}$^{-1}$ [Fig.~\ref{fig:INS_BMG}(b)], corresponding to the conventional first sharp diffraction peak of the glass state.
This is in agreement with the previous report~\cite{Komatsu2015}.
In the inelastic regime, a broad spectrum rising from $Q=2.6$~{\AA}$^{-1}$ is observed.
Its intensity enhances for larger momentum transfers, exhibiting an acoustic phonon-like $Q$ dependence governed by phonon polarization factors. 
This behavior indicates that the observed intensities likely originate from collective atomic vibrations.
A slight enhancement of the intensity below $Q=2$~{\AA}$^{-1}$ is possibly associated with atomic short-range correlations in the alloy. 
In the low-$Q$ regime, a broad peak is observed near 5~meV [see Fig.~\ref{fig:INS_BMG}(c)], which can be attributed to a boson peak arising from low-frequency vibrational modes commonly found in glassy materials~\cite{Buchenau1984,Sette1998,Li2008}.
Unlike the sharp phonon features in the commonly used alloys~\cite{Kibble2019}, which complicate the extraction of sample signals, the broad features in the Zr-BMG are expected to be advantageous for separating sample signals from background scattering in high-pressure INS experiments.

\subsection{Empty cell INS background}
\label{sec:empty}
\begin{figure}[!t]
\centering
\includegraphics[scale=1]{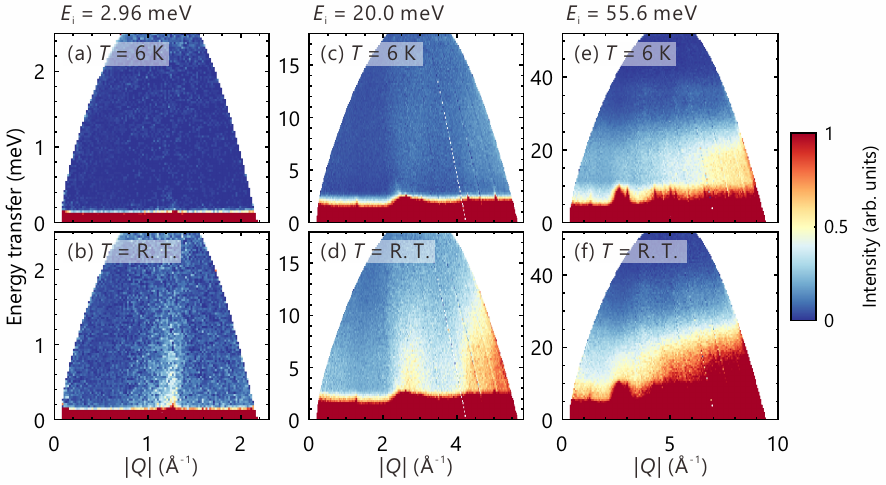}
\caption{False-color plots of the INS spectra of the empty pressure cell measured with (a),(b) $E_{\rm i}=2.96$~meV, (c),(d) $E_{\rm i}=20.0$~meV, and (e),(f) $E_{\rm i}=55.6$~meV. Data were taken at $T=6$~K (upper panels) and at room temperature (R.T., lower panels).}
\label{fig:INS_cell}
\end{figure}
\begin{figure}[!t]
\centering
\includegraphics[scale=1]{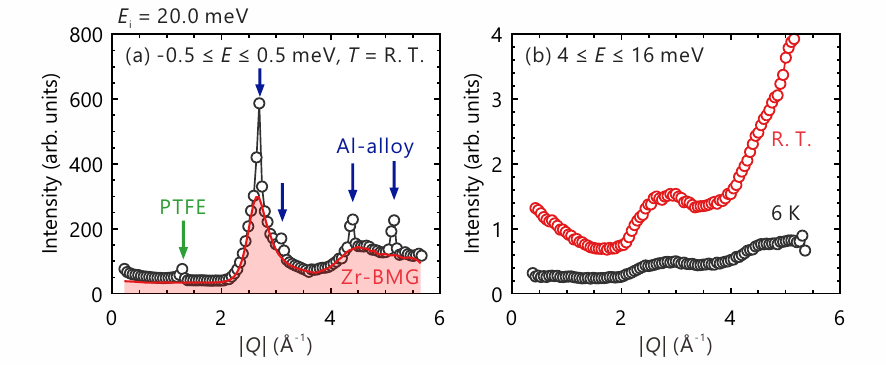}
\caption{(a) Elastic line of the INS spectrum of the pressure cell (black symbols) and the Zr-BMG rod (red curve), measured at R.T. with $E_{\rm i}=20.0$~meV. The energy transfers are integrated over $-0.5\leq E \leq 0.5$~meV. The intensity of the Zr-BMG rod is scaled by an arbitrary factor for visibility. Blue and green arrows indicate nuclear Bragg peaks arising from the aluminum-alloy body and the PTFE capsule, respectively. (b) Momentum-transfer profiles of the INS intensities of the cell integrated over $4\leq E \leq 16$~meV at $T=6$~K (black circles) and R.T. (red circles).}
\label{fig:INS_cell_1Dcut}
\end{figure}
INS spectra of the empty pressure cell, without a sample or pressure medium, were measured at 6~K and room temperature using several incident energies up to 55.6~meV.
Representative spectra are displayed in Fig.~\ref{fig:INS_cell}. 
A broad acoustic phonon-like spectrum, consistent with the INS spectrum of the Zr-BMG rod [Fig.~\ref{fig:INS_BMG}(a)], is observed at momentum transfers above $Q=2.6$~{\AA}$^{-1}$.
In addition, pronounced optical-phonon-like intensity appears around 20~meV, extending up to approximately 40~meV [Figs.~\ref{fig:INS_cell}(e) and \ref{fig:INS_cell}(f)].
A weak feature is also visible near $Q=1.15$~{\AA}$^{-1}$ at room temperature [Fig.~\ref{fig:INS_cell}(b)].

The elastic line data reveal nuclear Bragg peaks originating from the PTFE capsule~\cite{teflon} and the aluminum-alloy body, as indicated by the arrows in Fig.~\ref{fig:INS_cell_1Dcut}(a). 
Accordingly, the weak INS signal at $Q=1.15$~{\AA}$^{-1}$ [Fig.~\ref{fig:INS_cell}(b)] is attributed to the phonon intensity from the PTFE capsule.
Notably, the elastic intensities of the aluminum-alloy Bragg peaks are comparable to those of the Zr-BMG elastic scattering.
This probably reflects a strong suppression of the sharp phonon intensities typically observed for aluminum alloys~\cite{Kibble2019}.
Although the phonon-related features are visible in the inelastic regime at room temperature, they are significantly reduced upon cooling [Fig.~\ref{fig:INS_cell_1Dcut}(b)].
Consequently, the background from the Zr-BMG pressure cell is clean and featureless below $Q=2$~{\AA}$^{-1}$ at low energies ($<10$~meV), similar to that of other commonly used alloys such as CuBe and NiCrAl~\cite{Kibble2019}.
In the high-momentum regime ($>2$~{\AA}$^{-1}$), the background is broader than those of CuBe and NiCrAl, which is beneficial for the extraction of sample signals.

\subsection{Test measurement on \CFC}
\begin{figure}[!b]
\centering
\includegraphics[scale=1]{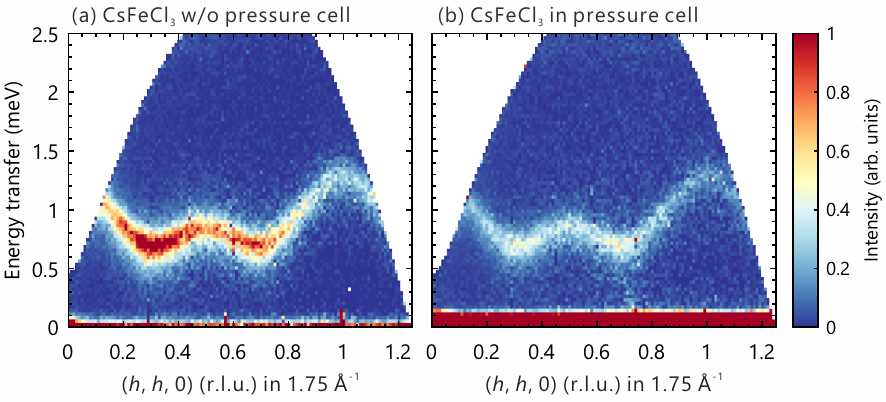}
\caption{False-color plots of the INS spectra of \CFC along the $[110]$ direction measured with $E_{\rm i}=2.96$~meV. Data were collected (a) without the pressure cell and (b) with the pressure cell. The momentum transfer perpendicular to the plot axis is integrated over $\pm0.1$~{\AA}$^{-1}$.}
\label{fig:INS_CsFeCl3_spectra}
\end{figure}

To evaluate the background quality and neutron transmission of the Zr-BMG pressure cell in the low-$Q$ ($<2$~{\AA}$^{-1}$) and low-energy ($<10$~meV) regime, we measured INS signals of \CFC in both the bare sample and the sample enclosed in the cell.
Magnetic excitations of the quantum magnet \CFC, known to exhibit a gapped dispersion relation~\cite{YoshizawaJPSJ1980}, are examined under the identical measurement conditions.
The INS intensities are normalized by the number of incident protons at the neutron source, enabling a direct and quantitative comparison between the two datasets.
As shown in Fig.~\ref{fig:INS_CsFeCl3_spectra}, the characteristic magnetic excitations, consistent with the previous studies~\cite{YoshizawaJPSJ1980,HayashidaPRB2019,HayashidaSciAdv2019,StoppelPRB2021}, were clearly observed in both the bare sample and the sample enclosed in the cell, even though the intensity is somewhat reduced in the latter.

\begin{figure}[!ht]
\centering
\includegraphics[scale=1]{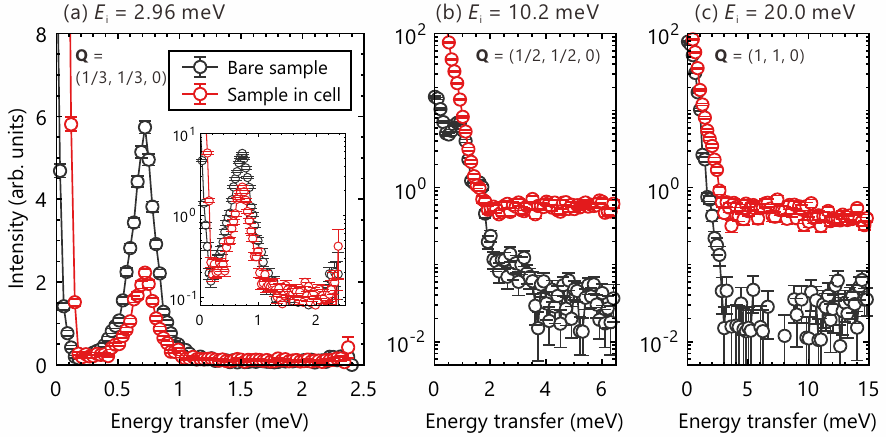}
\caption{Constant-$Q$ cuts of the INS spectra of \CFC measured without and with the pressure cell for (a) $E_{\rm i}=2.96$, (b) 10.2, and (c) 20.0~meV. The cuts are taken at (a) $\mathbf{Q}=(1/3,1/3,0)$, (b) $(1/2,1/2,0)$, and (c) $(1,1,0)$, each lying below 1.75~{\AA}$^{-1}$. For $E_{\rm i}=10.2$ and 20.0~meV, the $\mathbf{Q}$ positions are chosen to access higher energy transfers. Momentum transfers are integrated within $\pm0.1$~{\AA}$^{-1}$ along the three orthogonal directions. The inset of (a) and panels (b) and (c) are plotted on a logarithmic vertical scale.}
\label{fig:INS_CsFeCl3_1Dcut}
\end{figure}

The reduction of the magnetic signal is estimated from a constant-$Q$ cut at $\mathbf{Q}=(1/3,1/3,0)$, as shown in Fig.~\ref{fig:INS_CsFeCl3_1Dcut}(a).
The integrated INS intensity at 0.7~meV decreases to approximately 34{\%} of the bare-sample value.
To validate this value, we calculate the neutron transmission of the pressure cell materials.
The transmission $T$ is given by
\begin{equation}\label{eq:transmission}
T = \exp\left(-\frac{\rho N_{\rm A}}{M}\sigma_{\rm tot}t\right),
\end{equation}
where $\rho$ is the mass density of the material, $M$ is its molar mass, $N_{\rm A}$ is the Avogadro constant, and $t$ is the thickness of the material.
The total neutron cross-section $\sigma_{\rm tot}$ represents the sum of the cross sections of all the elements in the materials. 
The values of $\sigma_{\rm tot}$ for each element are taken from the JENDL-4.0 neutron cross-section data library~\cite{JENDL4}.
The calculated neutron-energy dependence of the transmission is shown in Appendix~\ref{sec:transmission}.
The neutron transmission values at 2.96~meV are summarized in Table~\ref{tb:neutron_transmission}, with the value for the monobloc CuBe cell~\cite{Hattori2022} provided for comparison.
For the present Zr-BMG cell, the transmission at the energy loss of 0.7~meV with the incoming and outgoing neutrons of 2.96 and 2.26~meV is estimated to be 33{\%}, in good agreement with the reduction extracted from the measured INS intensity.
Remarkably, the transmission of the Zr-BMG cell for this neutron energy is about 2.5 times larger than that of the CuBe piston-cylinder clamp cell.

\begin{table}
\tbl{Neutron transmission $(T)$ at a neutron energy of 2.96~meV for the Zr-BMG and aluminum alloy, together with their material parameters: mass density $\rho$, molar mass $M$, total neutron cross-section $\sigma_{\rm tot}$, and thickness $t$, as described in the text. The chemical composition of the aluminum alloy follow the Japanese Industrial Standard (JIS), and mass densities are given as representative values. The transmission of a monobloc CuBe cell with a thickness of 1.9~cm is taken from Ref.~\cite{Hattori2022}.}
{\begin{tabular}{lccccc} \toprule
Material & $\rho$ (g/cm$^{3}$) & $M$ (g/mol) & $\sigma_{\rm tot}$ (b) & $t$ (cm) & $T$ ({\%}) \\ \midrule
Zr$_{55}$Al$_{10}$Ni$_{5}$Cu$_{30}$ & 6.8 & 748.7 & 117.6 & 1.35 & 42 \\
Aluminum alloy (A7075) & 2.8 & 26.98 & 2.32 & 1.35 & 81 \\
CuBe monobloc~\cite{Hattori2022} & -  & - & - & 1.9 & 13 \\ \bottomrule
\end{tabular}}
\label{tb:neutron_transmission}
\end{table}

In addition to the superior transmission of the Zr-BMG cell, the Zr-BMG cell also provides a clean background in the INS spectra of samples in the cell.
In the low-energy regime, as shown in the inset of Fig.~\ref{fig:INS_CsFeCl3_1Dcut}(a), the background above 1.5~meV is nearly identical to that observed without the cell.
At the higher incident energies, the background increases by roughly two orders of magnitude [Figs.~\ref{fig:INS_CsFeCl3_1Dcut}(b) and \ref{fig:INS_CsFeCl3_1Dcut}(c)], yet remains featureless.
This background enhancement may be attributed to atomic short-range correlations of the alloys, the boson peak of the Zr-BMG, and incoherent scatterings from the atomic vibration within the pressure cell.
Importantly, the smooth and structureless background scattering likely enables straightforward background subtraction in INS data analysis.
Overall, the achievement of the high neutron transmission and clean background demonstrates that the Zr-BMG piston-cylinder clamp cell provides a clear advantage over conventional cells, such as those made of CuBe, for high-pressure INS experiments.

\section{Conclusions}
We have developed and characterized a Zr-BMG hybrid clamp cell designed for high-pressure INS experiments. 
The INS data of the empty cell exhibited broad and featureless spectra, yielding a clean background profile that enhances the visibility of sample signals. 
Test measurements using a reference sample \CFC demonstrated that the Zr-BMG cell provided significantly higher neutron transmission than a conventional monobloc CuBe clamp cell. 
These results establish Zr-BMG as a highly promising alternative pressure-cell material for high-pressure INS experiments.
For the future perspective, the present Zr-BMG pressure cell provides strong potential for investigating a wide range of quantum materials, including magnetic systems, quantum spin materials and unconventional superconductors.

\section*{Acknowledgements}
We thank the sample-environment team of J-PARC$\cdot$MLF for their technical supports on the cooling tests of the pressure cell. We also appreciate the assistance of K.~Namba and T.~Honda in the synthesis of \CFC crystals.
The synthesis of Zr-BMG was conducted under GIMRT user program (Project No.~202412-RDKGE-0040), organized by Institute for Materials Research, Tohoku University.
The magnetic property characterization of the synthesized Zr-BMG and the crystal growth of \CFC were carried out using a Quantum Design Magnetic Property Measurement System (MPMS) and a vertical tube furnace installed at the CROSS User Laboratory. 
The neutron experiments at the 4SEASONS spectrometer (J-PARC$\cdot$MLF) were carried out under Proposal No.~2025C0001 (CROSS Development Use), and Proposal Nos.~2024I0001 and 2025I0001 (Instrument Use).
Additional neutron experiment at HQR (JRR-3, JAEA) was conducted under the IRT program  (Proposal Nos.~25408 and 25412), organized by the Institute for Solid State Physics, The University of Tokyo.
This study was supported in part by JSPS KAKENHI:  JP23H04867 [Grant-in-Aid for Transformative Research Areas (A)] and JP24K00574 [Scientific Research (B)].

\appendix
\section{Pressure generation}
\label{sec:HQR}

\begin{figure}[!t]
\centering
\includegraphics[scale=1]{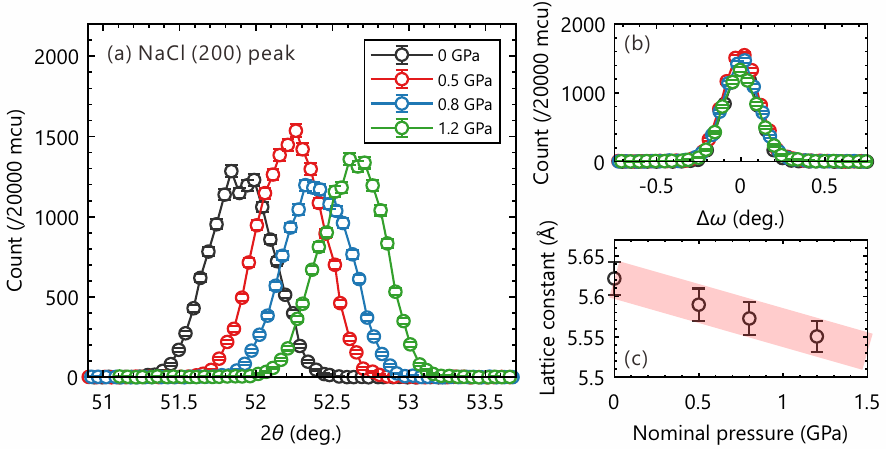}
\caption{(a) Pressure dependence of $\theta$–$2\theta$ scans around the NaCl $(200)$ Bragg peak. Neutron intensities are normalized to the monitor count unit (mcu). (b) $\omega$ scans around the $(200)$ peak at each pressure, with the peak centers nominally set to zero. (c) Pressure evolution of the NaCl lattice constant, estimated from the Gaussian function fitting of the $\theta$-$2\theta$ scans. Error bars are estimated from the standard deviation of the fitted Gaussian peaks. Pressures correspond to nominal values estimated from the applied load. The red thick line is a guide for the eye.}
\label{fig:NaCl}
\end{figure}

The generation of pressure was examined through the variation of the NaCl lattice constant obtained by neutron diffraction measurements.
Neutron diffraction measurements were conducted using the high-$Q$ resolution spectrometer (HQR) installed at the T1-1 beam port of the JRR-3 research reactor, Japan Atomic Energy Agency (JAEA), Japan. 
A neutron wavelength of 2.46~{\AA} was selected with a pyrolytic graphite monochromator, and the horizontal collimation sequence was Guide-40$^{\prime}$-20$^{\prime}$-40$^{\prime}$.
A NaCl crystal (Crystal Base Co., Ltd.) with dimensions of approximately $3.5\times3.5\times10$~mm$^{3}$ was placed in the pressure cell with deuterated glycerol as the pressure-transmitting medium. 
The crystal was aligned such that the crystallographic $ab$ plane was horizontal, and the pressure cell was mounted directly on the sample stage.

Figure~\ref{fig:NaCl}(a) shows the $\theta$-$2\theta$ scans around the NaCl $(200)$ Bragg peak  at various applied pressures.
With increasing pressure, the Bragg peak shifts systematically toward higher angles, reflecting the compression of the NaCl lattice.
Importantly, no discernible $\theta$-$2\theta$ peak broadening is observed, and the crystal mosaicity remains unchanged throughout the pressure range, as evidenced by the $\omega$ scans shown in Fig.~\ref{fig:NaCl}(b). 
These observations confirm that the applied pressure is essentially hydrostatic.
The lattice constants, extracted from the peak positions of the $\theta$-$2\theta$ scans, decrease linearly with pressure [see Fig.~\ref{fig:NaCl}(c)].
The relative volume compression of NaCl between ambient pressure and the nominal pressure of 1.2~GPa, $(V_{0}-V)/V_{0}$, is estimated to be 0.0377.
Using the equation of state $V(P,T)$ for NaCl~\cite{Brown1999}, the calibrated pressure is determined to be 1.0~GPa.
These results demonstrate that the Zr-BMG pressure cell provides reliable and homogeneous pressurization of the sample.

\section{Simulation of neutron transmissions}
\label{sec:transmission}

\begin{figure}[!t]
\centering
\includegraphics[scale=1]{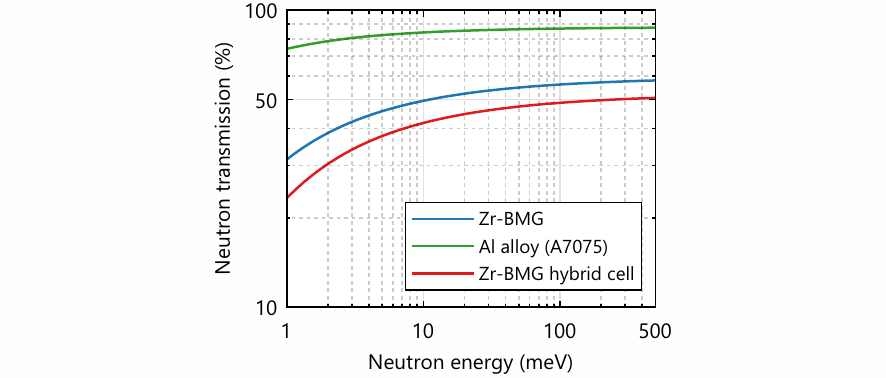}
\caption{Neutron energy dependence of neutron transmission for the Zr-BMG, aluminum alloy (A7075), and Zr-BMG hybrid cell. The material thicknesses are 1.35~cm for the Zr-BMG and aluminum alloy components. The transmission of the Zr-BMG hybrid cell is calculated by combining the transmissions of the Zr-BMG and aluminum alloy.}
\label{fig:transmission}
\end{figure}

Neutron transmission through materials generally depends on the energy of the neutrons. 
Figure~\ref{fig:transmission} shows the neutron-energy dependence of the transmission, calculated using Eq.~(\ref{eq:transmission}), for Zr-BMG, the aluminum alloy (A7075), and the Zr-BMG hybrid pressure cell developed in this study.
In the calculation, both incident and scattered neutrons are assumed to pass through the materials in a planar geometry, with a thickness of 1.35~cm for both the Zr-BMG and the aluminum alloy.
The energy dependence of the total cross-section $\sigma_{\rm tot}$ is taken from the JENDL-4.0 database~\cite{JENDL4}.
Note that these transmissions do not include additional effects such as Bragg edges or phonon contributions.
Notably, the Zr-BMG cell exhibits higher neutron transmission than the monobloc CuBe cell~\cite{Hattori2022} over the entire energy range considered.
In particular, its transmission exceeds that of the CuBe cell~\cite{Hattori2022} by more than a factor of two below 10~meV.

\bibliographystyle{tfnlm}
\bibliography{ZrBMG_INS}


\end{document}